\documentclass[11pt,reqno]{amsart}
\usepackage[letterpaper, portrait, margin=1in]{geometry}
\makeatletter
\g@addto@macro{\endabstract}{\@setabstract}

\usepackage{amsmath}
\usepackage{latexsym}
\usepackage{amsfonts}
\usepackage{amssymb}
\usepackage{lipsum}

\usepackage{bbm,dsfont}
\usepackage{dsfont}
\usepackage{graphicx}
\usepackage{hyperref}

\usepackage{verbatim}

\newtheorem{proposition?}{Proposition?}

\newtheorem{definition}{Definition}
\newcommand{\authorfootnotes}{\renewcommand\thefootnote{\@fnsymbol\c@footnote}}%
\makeatother

\usepackage{color}

\newcommand{\red}[1]{\textcolor{red}{#1}}
\newcommand{\green}[1]{\textcolor{green}{#1}}
\newcommand{\blue}[1]{\textcolor{blue}{#1}}
\newcommand{\cyan}[1]{\textcolor{cyan}{#1}}
\newcommand{\Ps}{\mathsf{P}}

\usepackage[normalem]{ulem} 

\newcommand{\away}[1]{\red{\sout{#1}}}
\newcommand{\instead}[1]{\blue{#1}}
\newcommand{\margincomment}[1]{{\color{blue}\rule[-0.5ex]{2pt}{2.5ex}}\marginpar{\small\begin{flushleft}\color{blue}#1\end{flushleft}}}

\newcommand{\msout}[1]{\text{\sout{\ensuremath{#1}}}}

\usepackage{graphicx}
\usepackage{xcolor}
\usepackage{graphics}
\usepackage{amsmath}
\usepackage{latexsym}
\usepackage{amsfonts}
\usepackage{amssymb}
\usepackage{lipsum}

\usepackage{bbm,dsfont}
\usepackage{dsfont}
\usepackage{graphicx}
\usepackage{hyperref}

\usepackage{verbatim}

\usepackage[normalem]{ulem} 

\newcommand{\po}{{\sc pom }}

\newcommand{\rational}{\mathbb Q} 
\newcommand{\real}{\mathbb R} 
\newcommand{\complex}{\mathbb C} 
\newcommand{\integer}{\mathbb Z} 
\newcommand{\half}{\tfrac{1}{2}} 
\newcommand{\mo}[1]{\left| #1 \right|} 
\newcommand{\mods}[1]{\left \vert #1 \right \vert ^2}
\newcommand*\colvec[3][]{
    \left(\begin{smallmatrix}\ifx\relax#1\relax\else#1\\\fi#2\\#3\end{smallmatrix}\right)
}

\newcommand{\hi}{\mathcal{H}} 
\newcommand{\his}{\mathcal{H}_{\mathcal{S}}}
\newcommand{\hir}{\mathcal{H}_{\mathcal{R}}}
\newcommand{\hia}{\mathcal{H}_{\mathcal{A}}}
\newcommand{\hik}{\mathcal{K}} 
\newcommand{\hv}{\mathcal{V}} 
\newcommand{\hr}{\mathcal{H}_{\mathcal{R}}}
\newcommand{\hs}{\mathcal{H}_{\mathcal{S}}}
\newcommand{\hit}{\mathcal{H}_{\mathcal{T}}}
\newcommand{\Sy}{\mathcal{S}}
\newcommand{\R}{\mathcal{R}}
\newcommand{\C}{\mathcal{C}}
\newcommand{\hic}{\mathcal{H}_{\mathcal{C}}}
\newcommand{\Y}{\yen}

\newcommand{\lh}{\mathcal{L(H)}} 
\newcommand{\lhs}{\mathcal{L}(\mathcal{H}_{\mathcal{S}})} 
\newcommand{\lht}{\mathcal{L}(\mathcal{H}_{\mathcal{T}})} 
\newcommand{\lha}{\mathcal{L}(\hia)}
\newcommand{\lhr}{\mathcal{L}(\hir)} 
\newcommand{\lk}{\mathcal{L(K)}} 

\newcommand{\linv}{\lh_{\text{sa}}^{\text{inv}}}

\newcommand{\bh}{\mathbf{B}(\mathcal{H})}
\newcommand{\bk}{\mathbf{B}(\mathcal{K})}
\newcommand{\trh}{\mathcal{T(H)}} 
\newcommand{\trk}{\mathcal{T(K)}} 
\newcommand{\sh}{\mathcal{S(H)}} 
\newcommand{\eh}{\mathcal{E(H)}} 
\newcommand{\ph}{\mathcal{P(H)}} 
\newcommand{\ip}[2]{\left\langle\,#1\,{|}\,#2\,\right\rangle} 
\newcommand{\ket}[1]{|#1\rangle} 
\newcommand{\bra}[1]{\langle#1|} 
\newcommand{\state}[1]{|#1\rangle} 
\newcommand{\dstate}[1]{\langle#1|} 
\newcommand{\kb}[2]{|#1\rangle\langle#2|} 
\newcommand{\no}[1]{\left\|#1\right\|} 
\newcommand{\nos}[1]{\left\|#1\right\|^2} 
\newcommand{\tr}[1]{\textrm{tr}\left[#1\right]} 
\newcommand{\ptr}[1]{\textrm{tr}_1[#1]} 
\newcommand{\pptr}[1]{\textrm{tr}_2[#1]} 
\newcommand{\trv}[1]{\textrm{tr}_\hv[#1]} 
\newcommand{\trvin}[1]{\textrm{tr}_{\hv_1}[#1]} 
\newcommand{\trvout}[1]{\textrm{tr}_{\hv_2}[#1]} 
\newcommand{\com}{\textrm{com}} 
\newcommand{\lb}[1]{lb(#1)} 
\newcommand{\ran}{\textrm{ran}\,} 
\newcommand{\id}{\mathbbm{1}} 
\newcommand{\nul}{O} 
\newcommand{\fii}{\varphi}
\newcommand{\Esf}{\mathsf{E}}
\newcommand{\Ecan}{\Esf ^{\text{can}}}
\newcommand{\Fsf}{\mathsf{F}}
\newcommand{\Zsf}{\mathsf{Z}}
\newcommand{\Gsf}{\mathsf{G}}
\newcommand{\Lsf}{\mathsf{L}}
\newcommand{\Msf}{\mathsf{M}}
\newcommand{\Asf}{\mathsf{A}}
\newcommand{\al}{\mathfrak{A}}
\newcommand{\Bsf}{\mathsf{B}}
\newcommand{\Ap}{\mathcal{A}}
\newcommand{\Bp}{\mathcal{B}}
\newcommand{\Rmc}{\mathcal{R}}
\newcommand{\Smc}{\mathcal{S}}
\newcommand{\Int}{\mathcal{I}_\epsilon}
\newcommand{\Ns}{N_\Smc}
\newcommand{\Nr}{N_\Rmc}
\newcommand{\M}{\mathcal{M}}
\renewcommand{\L}{\mathcal{L}}
\newcommand{\T}{\mathcal{T}}

\newcommand{\meo}{\Omega} 
\newcommand{\salg}{\mathcal{B}(\Omega)} 
\newcommand{\var}{\textrm{Var}} 
\newcommand{\bor}[1]{\mathcal{B}(#1)} 
\newcommand{\ltwo}[1]{L^2(#1)} 
\newcommand{\fidelity}{\mathcal{F}} 

\newcommand{\va}{\mathbf{a}} 
\newcommand{\vb}{\mathbf{b}} 
\newcommand{\vc}{\mathbf{c}} 
\newcommand{\ve}{\mathbf{e}} 
\newcommand{\vf}{\mathbf{f}} 
\newcommand{\vg}{\mathbf{g}} 
\newcommand{\vu}{\mathbf{u}} 
\newcommand{\vn}{\mathbf{n}} 
\newcommand{\vnn}{\hat\vn} 
\newcommand{\vm}{\mathbf{m}} 
\newcommand{\vk}{\mathbf{k}} 
\newcommand{\vx}{\mathbf{x}} 
\newcommand{\vy}{\mathbf{y}} 
\newcommand{\vsigma}{\boldsymbol{\sigma}} 
\newcommand{\vnull}{\mathbf{0}}
\newcommand{\gams}{ \gamma_{\theta}^{(\mathcal{S})}}
\newcommand{\gamr} {\gamma_{\theta}^{(\mathcal{R})}}

\newcommand{\Aa}{A(1,\va)} 
\newcommand{\Aan}{A(1,-\va)}
\newcommand{\Ab}{A(1,\vb)} 
\newcommand{\Abn}{A(1,-\vb)}
\newcommand{\An}{A(1,\vn)} 
\newcommand{\Ann}{A(1,-\vn)}
\newcommand{\Aaa}{A(\alpha,\va)} 
\newcommand{\Aaan}{A(2-\alpha,-\frac{\alpha}{2-\alpha}\va)}
\newcommand{\Abb}{A(\beta,\vb)} 
\newcommand{\Abbn}{A(1,-\vb)}
\newcommand{\Aoa}{A(1,\alpha\va)} 
\newcommand{\Aob}{A(1,\beta\vb)} 
\newcommand{\Qsf}{\mathsf{Q}}

\newcommand{\dev}{\mathfrak{D}}

\newcommand{\A}{\mathsf{A}}
\newcommand{\B}{\mathsf{B}}
\newcommand{\E}{\mathsf{E}}
\newcommand{\F}{\mathsf{F}}
\newcommand{\G}{\mathsf{G}}
\newcommand{\Q}{\mathsf{Q}}
\renewcommand{\P}{\mathsf{P}}

\newcommand{\Rmb}{\mathbb{R}}

\newcommand{\Ea}{E^{1,\va}} 
\newcommand{\Eb}{E^{1,\vb}} 
\newcommand{\Ec}{E^{1,\vc}} 
\newcommand{\Eaa}{E^{\alpha,\mathbf{a}}} 
\newcommand{\Ebb}{E^{\beta,\mathbf{b}}} 

\begin{document}


\title{Relative Quantum Time}

\maketitle
\begin{center}

  \normalsize
  \authorfootnotes
  Leon Loveridge\footnote{leon.d.loveridge@usn.no}\textsuperscript{1} and 
Takayuki Miyadera\footnote{miyadera@nucleng.kyoto-u.ac.jp}\textsuperscript{2}

  \textsuperscript{1}Department of Science and Industry Systems, University of South-Eastern Norway, Kongsberg, Norway  \par
  \textsuperscript{2}Department of Nuclear Engineering, Kyoto University, Nishikyo-ku, Kyoto, Japan 615-8540 \par
   \bigskip

\end{center}

\date{\today}

\begin{abstract}
The need for a time-shift invariant formulation of quantum theory arises from
fundamental symmetry principles as well as heuristic cosmological considerations. Such a description then leaves open the question of how to reconcile global invariance with the perception of change, locally. By introducing relative time observables, we are able to make rigorous the Page-Wootters conditional probability formalism to show how local Heisenberg evolution
is compatible with global invariance.
\end{abstract}
\section{Introduction}

A basic question in physics is how to reconcile fundamental symmetries with the perceived
asymmetry in the physical world. More precisely: under the postulate that all observed quantities are invariant under a relevant fundamental symmetry group, how can one explain the extraordinary effectiveness of the commonly used, very convenient description of physical phenomena in terms of non-invariant observables?

In quantum theory, for example, one describes position measurements very accurately in terms of the space-translation-covariant position observable, while it is obvious that operationally what we call ``the position" of a particle is its position relative to a reference object or frame. The \emph{relative position} is the translation-invariant fundamental quantity, but physicists routinely substitute \emph{absolute position} for it, with impunity. The resolution is found in the fact that it is generally possible to \emph{externalise} the (quantum) reference system, thereby ignoring its degrees of freedom, or effectively treating it as a classical reference frame. The work \cite{rel3} reviews the history and development of this solution, presents a formal framework for its rigorous formulation and a precise specification of the conditions under which such externalisation is possible.

Here we consider the analogous problem for \emph{time}: how can the time translation invariance, and hence stationarity, obeyed by a closed system, be reconciled with the observed non-stationary Schr\"odinger (or Heisenberg) time evolution displayed by (some of) its subsystems? An answer to this question was presented in a paper by Page and Wootters in 1983 \cite{pw1} in a cosmological context. The idea is that a subsystem identified as a quantum clock provides time readings in terms of the values of a suitable dynamical variable, conditional upon which the expectation values of another subsystem evolve in line with the Heisenberg equation of motion, all whilst maintaining the time-translation-invariance at level of the full system. While this idea appears natural, its implementation has been criticised in the literature. 

In \cite{kk1}, for example, Kucha\v{r} pointed out a mathematical subtlety in the Page-Wootters construction of invariant observables (Dirac observables) - 
they employed an integral of a time-evolved operator, 
the result of which is typically trivial. Indeed, 
for a one-particle system with a Hamiltonian $H= P$ (the momentum operator) the long-time integral of a spectral measure of 
the position operator $\Qsf(\Delta)$ becomes an operator 
proportional to the identity. Rephrased 
in the Schr\"{o}dinger picture, there is no time-invariant 
normal state. 

In this paper we offer a mathematically precise alternative to the Page-Wootters proposal, presenting a derivation of ``local" Heisenberg evolution under the constraint of global time translation invariance, using the methods developed in \cite{rel1,rel2,rel3}. 
The key observation is to replace the naive long time 
integral by \emph{relativisation}, introduced in previous work. Thus we can introduce 
well-defined non-trivial invariant observables. 
Much in the spirit of \cite{pw1}, we will proceed by studying a number of idealised scenarios, which allows us to highlight the conditions under which this free evolution law emerges.

\section{Time and Relative Time Observables}

\subsection{Absolute time observables}

Time appears as a parameter $t$ in the Schr\"{o}dinger (or Heisenberg) equation. 
It is therefore often
understood as a given ``classical parameter", whose interpretation is firmly rooted in classical physics and has no quantum description.  
Already at this level, some interesting and controversial discussions have appeared (e.g., \cite{bte,bte1,bte2,mte}).
However, examination of physically realistic scenarios shows that time must be represented quantum mechanically. The current time is inferred from systems behaving as ``clocks", which are physical 
objects in the world, and according to the universality of quantum theory, any physical system must have a quantum description if we shift the so-called Heisenberg cut so that the quantum system contains the clock. 

A concrete example follows from considering free-falling particles. Suppose we set $H_{\C}= {P_{\C}^2}/{2m} - Q_{\C}$ acting in  
$\mathcal{H}_{\C}:=L^2(\mathbb{R})$. The momentum operator $P_{\C}$ 
works as a hand of the clock. This operator $T_{\C}:=P_{\C}$ is conjugate to $H_{\C}$ and it satisfies 
\begin{eqnarray*}
e^{i H_{\C} t} T_{\C} e^{-i H_{\C} t} = T_{\C} + t \id.  
\end{eqnarray*} 
For later use, we may consider a one-particle system whose Hamiltonian is $H_{\C}=P_{\C}$. In that case, the position $Q_{\C}$ of a particle plays the role of the hand of a clock. 

A drawback of the above examples is the two-sided unboundedness of the Hamiltonians. They do not have a vacuum and are therefore ``too ideal", or unphysical.  
It has long been known that in quantum theory time does not, in general, admit an expression as a self-adjoint operator canonically conjugate to a lower-bounded Hamiltonian \cite{pauli}. 
The perspective that quantum observables are properly represented by positive operator
valued measures re-opens the possibility of having a quantum description of time 
\cite{lud1,hol1,bgl1} in formal analogy, for instance, to unsharp space-translation-covariant POVMs representing
 position observables subject to some intrinsic imprecision. 

Let us consider a (clock) system described by a Hilbert space 
$\mathcal{H}_{\C}$ with Hamiltonian $H_{\C}$ acting on $\mathcal{H}_{\C}$. We denote by 
$\mathcal{L}(\mathcal{H}_{\C})$ the set of all bounded operators 
on $\mathcal{H}_{\C}$. 
Rather than seeking a self-adjoint operator canonically conjugate to the
 Hamiltonian, one may rather demand covariance under time translations, that is, a positive-operator-valued measure (POVM) 
$\Esf_{\C} : \mathcal{B}(\mathbb{R}) \to 
\mathcal{L}(\mathcal{H}_{\C})$ for which
\begin{equation}\label{eq:cov}
 e^{iH_{\C}t}\Esf_{\C}(X)e^{-iH_{\C}t} = \Esf_{\C}(X-t);
\end{equation}
here $t \in \mathbb{R}$, $\mathcal{B}(\mathbb{R})$ denotes the Borel sets and $t \mapsto e^{iH_{\C}t}$ constitutes a unitary representation of the time translation group. 
We call a POVM satisfying (\ref{eq:cov}) an {\em absolute 
 time observable}. 
The operator $T_{\C}
:=\int _{\mathbb{R}} t \Esf_{\C}(dt)$ is symmetric, and in general not self-adjoint and admits no self-adjoint extension. $T_{\C}$ is self-adjoint exactly when $\Esf_{\C}$ 
is projection-valued, in which case the above integral expression corresponds to the familiar spectral
resolution of $T_{\C}$. 
Many examples of absolute time 
observables are given in \cite{BuschBook}.

\subsection{Relative time observables}

In this paper, we consider also \emph{relative (or relational)
 time observables}. In \cite{rel1,rel2,rel3}, 
we argued that genuinely observable quantities 
in a fully quantum setting 
are those which are invariant under the action
of some symmetry transformation. 
For instance, the absolute position operator 
$Q_{\C}$ of a particle 
implicitly assumes 
a classical reference frame external to the quantum system. 
Thus a more precise formulation must have a quantum 
description of the reference 
system. $Q_{\C}$ is obtained as a sort of approximation of 
a relative observable $Q_{\C} - Q_{\R}$, where $Q_{\R}$ is a position 
operator of a reference object, under a certain condition which enables 
the reference object to be regarded as classical. 
In \cite{rel1,rel2,rel3} it was observed in general that the ordinary absolute description functions
as an adequate shorthand for the true, relative description, when the absolute quantities are
understood not in reference to single systems, but to compound systems, with the suppressed system
playing the role of a reference. 

Let us recall an example of an absolute time observable. 
A clock system has a Hamiltonian $H_{\C}= P_{\C}$ and an absolute time 
observable, a ``clock hand", is the position of the particle $T_{\C}=Q_{\C}$. According to the above argument, the position $Q_{\C}$ itself, however, implicitly assumes the existence of a reference system and 
is not the most precise/fundamental description. 
Therefore $T_{\C}$ is not either; a reference system is required to give it precise meaning. 
In our clock example, the position of the clock hand becomes meaningful only relative to the clock face. 
This example indicates that as well as the position of a particle, time must be understood as a relative quantity. 
In the last section we put the Heisenberg cut just outside the clock
system. 
We now shift the Heisenberg cut further so that a reference system
 is also on the quantum side. 
We assume that there exists a one-parameter symmetry transformation on the composite system of a clock and its reference system. Any observable on the composite system is assumed to be invariant with respect to the transformation. 

Here, we therefore impose the time-shift invariance requirement
at the level of compound systems. We introduce 
\emph{clock} 
$\mathcal{C}$ and \emph{reference} $\mathcal{R}$, with associated spaces 
$\hic$ and $\hir$ respectively. 

We now construct relative time observables on $\hic \otimes \hir$, noting that these may 
in principle be defined for any compound system. Let $\Zsf:\mathcal{B}(\mathbb{R}) \to \mathcal{L}(\hic\otimes \hir)$ 
be a POVM.
Consider Hamiltonians $H_{\C}$ and $H_{\R}$ 
acting in (dense domains of) their respective spaces, defining the respective 
unitary groups $V_{\C}(t) = e^{-iH_{\C}t}$ and $V_{\R}(t) = e^{-iH_{\R}t }$. 
\begin{definition}\label{def:rtime}
$\Zsf$ is called a \emph{relative time observable} if:
\begin{enumerate}
\item $\left(V_{\C}(t) \otimes V_{\R}(t)\right)^* \Zsf (\Delta)\left(V_{\C}(t) \otimes V_{\R}(t)\right) = \Zsf (\Delta)$ for all $\Delta \in \mathcal{B}(\mathbb{R})$ (Invariance)
\item $V_{\C}(t)^* \Gamma_{\rho}(\Zsf (\Delta))V_{\C}(t) = \Gamma_{\rho}(\Zsf(\Delta - t))$ for all $\Delta \in \mathcal{B}(\mathbb{R})$ and $\rho \in \mathcal{S}(\hi_{\R})$ (Covariance), 
where $\Gamma_{\rho}: \mathcal{L}(\hic \otimes \hir) \to \mathcal{L}
(\hic)$ is a partial trace with respect to a state $\rho$. 
\end{enumerate}
\end{definition}
In other words, relative time observables are invariant at the composite level and covariant under restriction. We note that the invariance requirement pertains to Hamiltonians which are \emph{additive} over the composite system, i.e., we do not consider any dynamical coupling. The existence of relative time observables is established through
\emph{relativisation} \cite{rel3}.
Suppose that we have absolute time observables 
$\Esf_{\C}$ and $\Esf_{\R}$ acting on $\hic$ and $\hir$ respectively. 
A relativisation of some operator $A$ acting in $\hic$ with respect to 
 $\Esf_{\R}$ is defined by 
\begin{equation}
A \mapsto \Y(A) := \int_{\mathbb{R}} e^{-i H_{\C} t} 
A e^{i H_{\C} t} \otimes \Esf_{\R}(dt). 
\end{equation} 
In particular, $(\Y\circ \Esf_{\C})(X) := \Y(\Esf_{\C}(X))$
becomes 
\begin{equation}
(\Y\circ\Esf_{\C})(X) = \int_{\mathbb{R}} \Esf_{\C}(X + t)\otimes \Esf_{\R}(dt). 
\end{equation} 
This quantity 
is invariant, given that $\Esf_{\R}$ is covariant, and the covariance of $(\Gamma_{\rho}\circ \Y) (\Esf_{\C})$ for all $\rho \in \mathcal{S}(\hi_{\R})$ follows from a simple calculation. 
In addition we may note that this can be rewritten as 
\begin{eqnarray}
(\Y\circ \Esf_{\C})(X)= \int_{\mathbb{R}} \Esf_{\C}(du) \otimes \Esf_{\R}(u -X),
\label{exchange} 
\end{eqnarray}
which implies that the relativisation is essentially same with 
the relativisation of $\Esf_{\R}$ with respect to $\Esf_{\C}$,
except for the unimportant sign. 

A concrete example follows from considering free-falling particles. Suppose we set $H_{\C}= {P_{\C}^2}/{2m} - Q_{\C}$ and $H_{\R}={P_{\R}^2}/{2m}+Q_{\R}$, both acting in (separate copies of) $L^2(\mathbb{R})$. It can be readily verified that a relational
time observable for $\C + \R$ is provided by the total momentum: the spectral measure $\Esf^{P}$ defined by the self-adjoint operator $P = P_{\C} + P_{\R}$ is manifestly invariant due to the 
differing signs on the potential terms in the total Hamiltonian, and the covariance of the restriction follows from the additivity of $P$. 


\section{Recovering the equation of motion}

\subsection{Conditional probability formalism}
In the last section, we introduced relative time observables $\Zsf$ which are regarded as genuine quantum descriptions of time. 
For this new description to be valid, there should be a regime in which we can regain the normal description of time as an external parameter. 
In the normal description, observables evolve, as time elapses, according to the Heisenberg equation of motion. 
Suppose that we have a system described by a Hilbert space $\his$ 
with Hamiltonian $H_{\Sy}$. 
Then the normal description claims that each operator $A$ 
evolves in time as $\alpha_t^{\Sy}(A):=
e^{i H_{\Sy} t} Ae^{-i H_{\Sy} t}$. 
The purpose of this section is to show how this equation of motion 
is recovered in our formalism in which all the observables are 
invariant with respect to time shifts, and thus apparently nothing evolves.

A key observation, inspired by \cite{pw1}, is to use the formalism of conditional probabilities. 
In realistic physical situations, when we claim that at time $t$ an observable $\A$ shows some value $x$, we measure both a clock and the observable. 
Therefore a more precise description of this statement is 
``when we observe a clock and obtain a value $t$, we obtain $x$ as a result of 
measuring $\A$''. Thus it needs conditioning on time. 
In the following we study two examples employing such a conditional 
probability statement to examine the relative time formalism.   

\subsection{Discrete Time}

The definitions in the previous sections are naturally extended 
to discrete periodic 
absolute and relative time observables by 
replacing $\mathbb{R}$ by $\mathbb{Z}_d$.  
We construct a model where both the clock and reference have discrete periodic (and sharp) time observables. 
Ordinary clocks have only $12 \times 60 \times 60$ seconds 
to be distinguished, and thus it is in a sense realistic. 
These are represented as the cyclic time in $\mathbb{C}^d$, with 
eigenstates 
$\ket{n}$ and eigenvalues $n=0,1,\dots,d-1$ counted cyclically, i.e., understood as 
elements of $\mathbb{Z}_d$. Then the self-adjoint absolute 
time operator is $T_{\C}=\sum n\ket{n}\bra{n} \equiv Q_{\C}$. 
In addition to the clock and reference, there is a system $\Sy$
in which we are interested, whose Hamiltonian is denoted by $H_{\Sy}$. 
It defines an  
action of the shift group ($k\in \mathbb{Z}_d$), 
given by 
$\alpha_k^{\Sy}(A) = e^{i H_{\Sy} k } A e^{-i H_{\Sy} k}$ 
for $\Sy$. Note that while 
we treat three systems and call 
the second and the third system a clock and a reference system, 
their names can be exchanged (see (\ref{exchange})). 

Let the total Hamiltonian be of the form
\[
H=H_{\Sy}+P_{\C}+P_{\R}.
\]
Here, e.g., $P_{\C}$ is the shift generating ``momentum" operator, $P=\sum m\ket{f_m}\bra{f_m}$, with $m\in \mathbb{Z}_d$ and $\ket{f_m}=
\frac{1}{\sqrt{d}}
\sum_ne^{2\pi i mn/d}\ket{n}$.
It defines an action $\alpha^{\C}_{k}(|n\rangle \langle m|)
=e^{i P_{\C} k } |n \rangle \langle m | e^{-i P_{\C} k} 
= |n-k\rangle \langle m-k|$. An action on the reference system 
is $\alpha^{\R}_k (|n\rangle \langle m | )
= e^{i P_{\R} k }|n \rangle \langle m | e^{-i P_{\R} k} 
= |n -k \rangle \langle m - k|$. 
Note that $\{|n\rangle \langle n |\}$ on each space is 
an absolute time observable.  
Any relative/relational observable must be invariant with respect to 
this total Hamiltonian. 
A relative time observable is obtained by relativising 
a POVM $\{|n\rangle \langle n|\} \subset \mathcal{L}(\hic)$ as, 
\begin{eqnarray*}
\Y(\id \otimes 
|n\rangle \langle n | ) = \sum_m \id \otimes |n +m \rangle \langle 
n+m | 
\otimes |m \rangle \langle m |. 
\end{eqnarray*}

Now let us consider a POVM $\A = \{\A(k)\}_k$ on the system,
which is an absolute observable we are interested in. 
As its relativised object with respect to 
the absolute time observable in the reference system, we introduce 
\begin{eqnarray*}
\Y(\A(k)\otimes \id)= \sum_m \alpha^{\Sy}_{-m} (\A(k)) \otimes 
\id \otimes |m\rangle \langle m |. 
\end{eqnarray*}

To study conditional probability, we have to introduce a joint measurement of relational observables 
$\{ \Y(\id \otimes |n\rangle \langle n|)\}$ and $\{\Y(\A(k)
\otimes \id)\}$. 
Since they commute with each other, they are jointly measurable. Moreover, since $\{\Y(\id \otimes 
|n\rangle \langle n|)\}$ is sharp, their jointly 
measuring observable is uniquely determined 
\cite{BuschBook}
as 
\[
M(k,n)=\sum_m\alpha_{-m}^\Sy(\Asf(k))\otimes\ket{n+m}\bra{n+m}\otimes\ket{m}\bra{m}.
\]
To examine the joint probability, we assume the total state is 
\[
\rho=\ket{\Psi}\bra{\Psi}=\ket{\psi^\Sy}\bra{\psi^\Sy}\otimes \ket{0}\bra{0}\otimes\ket{\xi}\bra{\xi}.
\]
Then the expectation value (probability) is
\begin{align}\label{eq:prob}
P(k,n)&=\sum_m\left\langle{\psi^\Sy}\big|\alpha_{-m}^\Sy(\Asf(k) \big|\psi^\Sy\right\rangle\,\big|\langle{0}\ket{n+m}|^2\,|\langle{m}\ket{\xi}|^2\nonumber\\
&=\left\langle{\psi^\Sy}\big|\alpha_{n}^\Sy(\Asf(k) \big|\psi^\Sy\right\rangle\,|\langle{-n}\ket{\xi}|^2.
\end{align}
As its marginal probability for time, we obtain
\[
P(n)=\sum_k P(k,n)=|\langle{-n}\ket{\xi}|^2.
\]
Assume these probabilities all to be non-zero, then the conditional probability becomes
\[
P(k|n)=\left\langle{\psi^\Sy}\big|\alpha_{n}^\Sy(\Asf(k) )\big|\psi^\Sy\right\rangle.
\]
This is the expectation of the `Heisenberg-evolved' observable $\Asf$. Several remarks are in order. First, we observe that this result holds for arbitrary $\Asf$. Second, it is of course
crucial that the expression $|\langle{n}\ket{\xi}|^2$ is non-vanishing for all $n \in \mathbb{Z}_d$, which demands that $\ket{\xi}$ is broadly spread out in time. The simplest choice
for such a state is $\ket{\xi} = \ket{f_m}$ for some $m$, i.e., an eigenstate of the reference
Hamiltonian. It is thus an invariant state.

We also observe that the state $\ket{\Psi}$ is unentangled. We may also consider the entangled state
\[
\ket{\Psi ^{\prime}}=\sum_\ell\lambda_\ell\ket{\varphi_\ell}\otimes \ket\ell\otimes\ket{\xi_\ell}, 
\]
and compute
\begin{align*}
P'(k,n)&=\sum_{m,\ell,\ell'}\lambda_\ell\overline{\lambda}_{\ell'}\langle\varphi_{\ell'}|\alpha_{-m}^\Sy(\Asf(k))|\varphi_{\ell}\rangle\,
\langle{\ell'}\ket{n+m}\bra{n+m}\ell\rangle\,\bra{\xi_{\ell'}}m\rangle\langle m\ket{\xi_\ell}\\
&=\sum_m|\lambda_{n+m}|^2\bra{\varphi_{n+m}}\alpha_{-m}^\Sy(\Asf(k))\ket{\varphi_{n+m}}\,|\bra{m}\xi_{n+m}\rangle|^2.
\end{align*}
With the choices 
\begin{equation*}
|\varphi_{l}\rangle = e^{-i H_{\Sy} l } |\psi^{\Sy}\rangle ~ \text{and}~
|\xi_l\rangle = e^{i H_{\R} l } |\xi\rangle, 
\end{equation*}
one obtains (noting that $\sum_m|\lambda_m|^2=1$)
\[
P'(k,n)=P(k,n).
\]
Thus the same distributions can be obtained 
also in this entangled state. 
However, as shown above, entanglement is not 
necessary in our argument. Because normally a clock and 
a reference system are macroscopic systems and they are 
spatially separated, we think the product state is easy to be 
realized and 
more 
reasonable. The possibility of achieving this result using unentangled states
is of independent interest, given claims in the literature that entanglement is responsible for subsystem quantum dynamics (e.g., \cite{glm1},\cite{mv1}) which would now seem 
to require further scrutiny.\footnote{It is worth pointing out also that the state in the Page-Wootters spin model is also unentangled.}  


\subsection{Continuous Time}

Let us consider a system $\his$, a clock $\hi_{\C}$ and a reference frame $\hi_{\R}$,
with $\hi{_\C} \simeq \hi_{\R} \simeq L^2(\mathbb{R})$. A model Hamiltonian for the combined system is provided as a direct generalisation of the Hamiltonian for the discrete time model, namely
\begin{equation*}
H= H_{\Sy} + P_{\C}+P_{\R} ,
\end{equation*}
again with $H_{\Sy}$ an arbitrary Hamiltonian of the system $\Sy$. Suppose we fix a state $\rho_{\C}$ of $\C$, which for simplicity we presume to be pure, and localised around the origin with respect 
to the position. Hence $\rho_{\C} = |\psi_{\C}\rangle 
\langle \psi_{\C}|$
with ${\rm supp}( \psi_{\C}) \subset [-\epsilon, \epsilon]$, and $\epsilon >0$. The combined state is then of the form $\rho=\rho_{\Sy}\otimes \rho_{\C} \otimes \rho_{\R}$.

Now let $\Qsf_{\C}$ and $\Qsf_{\R}$ denote the spectral measures of the position operators 
$Q_{\C}$ and $Q_{\R}$, which respectively satisfy the following covariance conditions:
\begin{equation*}
e^{i P_{\C} t}\Qsf_{\C}(\Delta) e^{-i P_{\C} t} = \Qsf_{\C}(\Delta-t)
\end{equation*}  
and
\begin{equation*}
e^{i P_{\R} t}\Qsf_{\R}(\Delta) e^{-i P_{\R} t} = \Qsf_{\R}(\Delta -t).
\end{equation*}
Relativizing $\Qsf_{\C}$ with respect to a covariant POVM 
$\Qsf_{\R}$ we obtain a relative time observable:
\begin{eqnarray*}
\Zsf(\Delta) := \int \Qsf_{\C}(\Delta+t) \otimes \Qsf_{\R}(dt).
\end{eqnarray*}
It is nothing but a spectral decomposition of 
a relative position observable  
$Q_{\C} - Q_{\R}$. 

Take a discrete POVM $\A=\{\A(k)\}$ of $\Sy$.
Its relativisation with respect to $\Qsf_{\R}$ is written as 
\begin{eqnarray*}
\Y(\A(k))=
\int e^{-i H_{\Sy} t} \A(k) e^{i H_{\Sy} t} \otimes \Qsf_{\R}(dt). 
\end{eqnarray*}
Now we consider a joint measurement of the relative 
time $\Zsf$ and a relative observable $\Y(\A(k))$. 
As $\Zsf$ is a sharp observable, their jointly measuring observable 
is uniquely determined as, 
\begin{eqnarray*}
M(k, \Delta):= 
\int e^{-iH_{\Sy} t} \A(k) e^{i H_{\Sy} t} \otimes \Qsf_{\C}
(\Delta +t) \otimes \Qsf_\R(dt),
\end{eqnarray*} 
which is invariant under time translation. The expectation of $M(n, \Delta)$ in the state $\rho$ is
\begin{eqnarray*}
\langle M(k, \Delta) \rangle_{\rho} := \mbox{tr}[\rho M(k, \Delta)]
= \int \mbox{tr}[\rho_{\Sy} e^{-i H_{\Sy} t} \A(k)
 e^{i H_{\Sy} t}]
\langle \psi_{\C} | \Qsf_{\C}(\Delta +t)
|\psi_{\C}\rangle \mbox{tr}[\rho_{\R} \Qsf_{\R}(dt)]. 
\end{eqnarray*}
Informally putting $\mbox{tr}[\rho_{\R} \Q_{\R}(dt)] = f_{\R}(t) dt$ (which is justified due to the absolute continuity of $X \mapsto \mbox{tr}[\rho_{\R} \Q_{\R}(X)]$), and setting $\Delta= 
[t_0-\delta, t_0+\delta]$, we obtain 
\begin{eqnarray*}
\langle M(k, \Delta) \rangle_{\rho}
&=& \int \mbox{tr}[\rho_{\Sy} e^{-i H_{\Sy} t} \A(k)
 e^{i H_{\Sy} t}]
\langle \psi_{\C} | \Qsf_\C(\Delta +t)|\psi_{\C}\rangle f_R(t) dt
\\
&=&
\int^{-t_0+\delta +\epsilon}_{-t_0-\delta -\epsilon}
\mbox{tr}[\rho_{\Sy} e^{-i H_{\Sy} t} \A(k) e^{i H_{\Sy} t}]
\langle \psi_{\C}
 |\Qsf_\C(\Delta+t) |\psi_{\C} \rangle f_R(t) dt
\\
&=&
\int^{t_0 + \delta + \epsilon}_{t_0 - \delta -\epsilon}
\mbox{tr}[\rho_{\Sy} e^{i H_{\Sy} t} \A(k) e^{-i H_{\Sy} t}]
\langle \psi_{\C} |\Qsf_\C(\Delta -t)|\psi_{\C}\rangle
f_{\R}(-t) dt,  
\end{eqnarray*}
where we used the support property of $\psi_{\C}$, 
and 
\begin{eqnarray*}
\sum_k \langle M(k, \Delta)\rangle_{\rho}
= \int^{t_0 + \delta + \epsilon}_{t_0 - \delta -\epsilon}
\langle \psi_\C | \Qsf_\C(\Delta -t) |\psi_\C\rangle f_R(-t) dt, 
\end{eqnarray*}
which does not vanish for broadly extended $f_R(\cdot)$. 
Thus we obtain a conditional probability 
\begin{eqnarray*}
P(k|[t_0 - \delta, t_0 + \delta])
&=&  \frac{
\int^{t_0+\delta +\epsilon}_{t_0-\delta -\epsilon}
\mbox{tr}[\rho_{\Sy} e^{i H_{\Sy} t} \A(k) e^{-i H_{\Sy} t}]
\langle \psi_{\C} |\Qsf_{\C}(\Delta-t) |\psi_\C \rangle f_{\R}(-t) dt}
{\int^{t_0 + \delta + \epsilon}_{t_0 - \delta -\epsilon}
\langle \psi_\C | \Qsf_C(\Delta -t) |\psi_\C\rangle f_R(-t) dt
}
\\
&\simeq& \mbox{tr}[\rho_{\Sy} e^{i H_{\Sy} t_0} \A(k) e^{- iH_{\Sy} t_0}] 
\end{eqnarray*}
for sufficiently broad $f_{\R}$ and small $\delta, \epsilon$. 
It is nothing but the Heisenberg equation of motion. 

To study the quality of approximation, it is useful to introduce the characteristic function 
$\chi_{\Delta}(\cdot)$ (and to replace it by a general function $h$) and take the
Fourier transform. 
Let us examine the limit procedure in the Fourier transformed form. 
We introduce a smooth positive function $h(\cdot)$ 
which has a compact support and satisfies $0 \leq h(s) \leq 1$.
It defines an effect $\int h(s) \Zsf(ds)$ whose ``click'' means 
that the clock shows time in the support of $h$.   
Instead of $M(n, \Delta)$, we consider 
\begin{eqnarray*}
M(k, h) &:=& \int h(s) \alpha^{\Sy}_{-t} (\A(k)
) \otimes \Qsf_\C(ds +t) \otimes \Qsf_\R(dt)
\\
&=& \int h(\tau-t) \alpha^{\Sy}_{-t}(\A(k))
\otimes \Qsf_\C(d\tau) \otimes \Qsf_\R(dt). 
\end{eqnarray*}
Putting $h(s)=\chi_{\Delta}(s)$, we regain the original $M(n, \Delta)$.
We again introduce a function $f_{\C}$ formally by 
$f_{\C}(\tau) d\tau = \mbox{tr}[\rho_{\C} \Qsf_{\C}(d\tau)]$.
The conditional probability is written as 
\begin{eqnarray*}
\frac{\langle M(k, h)\rangle_{\rho}}{\langle E(h)\rangle_{\rho}}
=\frac{\int d\tau \int dt h(\tau-t) 
\mbox{tr}[
\rho_{\Sy}
 \alpha^{\Sy}_{-t}(\A(k))] f_\C(\tau) f_\R(t)}
{\int d\tau \int dt h(\tau-t) f_\C(\tau) f_\R(t)}. 
\end{eqnarray*}
In the energy representation, $\rho_{\Sy}$ is written as 
$\rho_{\Sy} = \sum \sum |\epsilon_m \rangle \langle \epsilon_m |\rho_{\Sy} |\epsilon_n\rangle 
\langle \epsilon_n |$.

Thus the conditional probability is written as 
\begin{eqnarray*}
\frac{\langle M(k, h)\rangle_{\rho}}{\langle E(h)\rangle_{\rho}}
=
\frac{\sum \langle \epsilon_m |\rho_{\Sy}|\epsilon_n \rangle
\langle \epsilon_n |\A(k)| \epsilon_m\rangle 
\int d \omega \tilde{f}_C(\omega) \tilde{f}_R(\epsilon_m-\epsilon_n -\omega) 
\tilde{h}(-\omega)
}
{\sum
\langle \epsilon_n |\rho_{\Sy}|\epsilon_n\rangle 
\int d\omega \tilde{f}_C(\omega) \tilde{f}_R(-\omega) \tilde{h}(-\omega) 
}, 
\end{eqnarray*}
where $\tilde{f}$ is defined by 
$\tilde{f}(\omega) = \frac{1}{\sqrt{2\pi}}\int f(t) e^{i\omega t}$.

Let us introduce a time-displaced $h$ by 
$h_{s}(t) = h(t-s)$.  
Its Fourier transform becomes 
$\tilde{h}_{s}(\omega)= e^{i \omega s}\tilde{h}(\omega)$. 
Thus we have 
\begin{eqnarray*}
\frac{\langle M(k, h_{s})\rangle_{\rho}}
{\langle E(h_{s})\rangle_{\rho} }
= 
\frac{\sum \langle \epsilon_m |\rho_{\Sy}|\epsilon_n \rangle
\langle \epsilon_n |\A(k)| \epsilon_m\rangle 
\int d \omega e^{-i \omega s} 
\tilde{f}_C(\omega) \tilde{f}_R(\epsilon_m-\epsilon_n -\omega) 
\tilde{h}(-\omega)
}
{
\int d\omega e^{-i \omega s} 
\tilde{f}_C(\omega) \tilde{f}_R(-\omega) \tilde{h}(-\omega) 
}.
\end{eqnarray*}
Let us control the broadness of $f_R$ by introducing a parameter $\lambda$
as 
\begin{eqnarray*}
f_R^{\lambda}(t):=\frac{1}{\lambda}f_R(t/\lambda).
\end{eqnarray*}
Then its Fourier transform becomes 
$\tilde{f}_R^{\lambda}(\omega)= \tilde{f}_R(\lambda \omega)$. 
Thus for reference states parametrized by $\lambda$, we have 
\begin{eqnarray*}
\frac{\langle M(k, h_{s})\rangle_{\lambda}}
{\langle E(h_{s})\rangle_{\lambda}}
= 
\frac{\sum \langle \epsilon_m |\rho_{\Sy}|\epsilon_n \rangle
\langle \epsilon_n |\A(k)| \epsilon_m\rangle 
\int d \omega e^{-i \omega s} 
\tilde{f}_C(\omega) \tilde{f}_R(\lambda(\epsilon_m-\epsilon_n -\omega)) 
\tilde{h}(-\omega)
}
{
\int d\omega e^{-i \omega s} 
\tilde{f}_C(\omega) \tilde{f}_R(-\lambda \omega) \tilde{h}(-\omega) 
}.
\end{eqnarray*}
One can see by changing variables properly that  for large $\lambda$ this converges to 
\begin{eqnarray*}
\lim_{\lambda \to \infty}
\frac{\langle M(k, h_{s})\rangle_{\lambda}}
{\langle E(h_{s})\rangle_{\lambda}}
&=&
\frac{
 \sum_{m,n}  \langle 
 \epsilon_m |\rho_{\Sy}|\epsilon_n \rangle
\langle \epsilon_n |\A(k)| \epsilon_m\rangle 
e^{-i (\epsilon_m -\epsilon_n) s}
\tilde{h}(-(\epsilon_m -\epsilon_n))\tilde{f}_{\C}(\epsilon_m - \epsilon_n)
}
{\tilde{h}(0)\tilde{f}_{\C}(0)}
\\
&=& 
\mbox{tr}[\rho_{\Sy} e^{i H_{\Sy}s}\overline{\A(k)}_{h, f_{\C}} e^{-i H_{\Sy}s}], 
\end{eqnarray*}
where $\overline{\A(k)}_{h, f_{\C}}$ is defined by 
\begin{eqnarray*}
\overline{\A(k)}_{h, f_{\C}} := 
\sum |\epsilon_n \rangle \langle \epsilon_n |\A(k) |\epsilon_m \rangle 
\langle \epsilon_m | \tilde{f}_{\C}(\epsilon_m - \epsilon_n ) \tilde{h}
(- (\epsilon_m - \epsilon_n))/ \tilde{f}_{\C}(0) \tilde{h}(0).
\end{eqnarray*}
Again in the limit of narrow support of $h$, it converges to 
\begin{eqnarray*}
\overline{\A(k)}_{h, f_{\C}} \to \overline{\A(k)}_{f_{\C}}
= \sum |\epsilon_n \rangle \langle \epsilon_n |\A(k) |\epsilon_m \rangle 
\langle \epsilon_m | \tilde{f}_{\C}(\epsilon_m - \epsilon_n )/
\tilde{f}_{\C}(0). 
\end{eqnarray*}
Thus we found that in the limit of broadly extended 
reference state the Heisenberg equation for an effective 
operator $\overline{\A(k)}_{f_{\C}}$ is recovered. 
This $\overline{\A(k)}_{f_{\C}}$ has a cutoff in the high-frequency part 
depending on the sharpness of the clock state. 

This can be interpreted in terms of \cite{rel1,rel3}. 
A measurement of the relative time observable 
essentially reduces a state of the reference system to a 
localized one with unsharpness of the clock state. 
We then measure a relativised observable of an 
absolute observable in the system. It was shown 
in \cite{rel1,rel3} that for this result 
to be close to 
the ideal one the unsharpness of the reference state is 
required to be small. 
\section{Discussion}

In this paper we introduced a formulation, 
an extension of the Page-Wootters formalism, 
which shows 
how dynamics emerges out of a ``frozen'', time invariant theory. 
Two observations played crucial roles. One is the introduction of 
a relative time observable, which shows essentially 
a ``difference'' between 
absolute time observables in a clock and a reference system. 
The relative observable is invariant and is covariant with respect to 
the time translation on the clock. Another is a formulation of the theory based on conditional probabilities. It naturally made us treat 
a joint measurement of the relative time observable and a relativized 
system observable. We examined two simple examples to show 
that our formulation recovers the ordinary Heisenberg equation of motion.  
In both discrete time and continuous time examples, we needed 
broadness (large uncertainty) in reference system states. Therefore  
the state on the reference system close to an energy eigenstate 
(or mixtures thereof) is found to work. In addition, in the continuous 
time example, we showed that a sharp clock state with respect to 
an absolute time observable is preferable. Its unsharpness introduces 
high-frequency cut-off effective observables. As mentioned, contrary to 
some existing formulations, our theory does not need any entanglement 
among the systems. Thus it works also in the classical theory. 
As maintaining entanglement among systems is difficult task, and 
normally our clock is a macroscopic object, we think that the irrelevance of the entanglement is reasonable. 
 
 Still there remain some issues to be addressed in our proposal. 
In addition to the subtlety of the definition of Dirac observables, 
Kucha\v{r} \cite{kk1} has pointed out that Page-Wootters' formulation 
gives incorrect propagators (see, however, \cite{glm1} for 
a recent proposal). A naive 
application of the sequential measurement machinery 
seems to show that our model also suffers from this issue. 
We think, however, that our model in a certain limit 
may give another 
conditional probability formulation proposed by Gambini et al. \cite{Gambini}, which overcomes such criticisms. 
We hope to address the problem elsewhere. 

\section*{Acknowledgements}
This work was part of an ongoing collaboration with Paul Busch, who died before the
manuscript could be finished. Paul had been interested in time in quantum theory 
since at least 1990 when he wrote his first papers \cite{bte1,bte2} on the time-energy uncertainty relation,
and his interest in invariance and the relative/absolute distinction had been ongoing since his work with LL during the latter's
PhD, starting in 2008. This topic brought Paul, LL and TM together in 2011 in York, and again in Kyoto
in 2018. Parts of the paper were written by Paul; LL and TM have made only minor edits where that is the case. This paper is dedicated to Paul, whose knowledge, insight and wisdom will be so sorely missed, and for those lucky enough to know him, his friendship too. We also thank Oliver Reardon-Smith for many helpful discussions
on the work presented here. This work was supported by JSPS KAKENHI Grant Number 15K04998.





\end{document}